# Vanadium-Doped Molybdenum Disulfide Monolayers with Tunable Electronic and Magnetic Properties: Do Vanadium-Vacancy Pairs Matter?

Da Zhou[1,2#], Yen Thi Hai Pham[3#], Diem Thi-Xuan Dang[3#], David Sanchez[4], Aaryan Oberoi[5], Ke Wang[6], Andres Fest[4], Alexander Sredenschek[1,2], Mingzu Liu[1,2], Humberto Terrones[7], Saptarshi Das[5], Dai-Nam Le[3], Lilia M. Woods[3], Manh-Huong Phan[3,*], and Mauricio Terrones[1,2,4,8,*]

[1]Department of Physics, The Pennsylvania State University, University Park, PA 16802

[2]Center for 2- Dimensional and Layered Materials, The Pennsylvania State University, University Park, PA 16802

[3]Department of Physics, University of South Florida, Tampa, FL 33620

[4]Department of Materials Science and Engineering, The Pennsylvania State University, University Park, PA 16802

[5]Department of Engineering Science and Mechanics, The Pennsylvania State University, University Park, PA 16802

[6]Materials Research Institute, The Pennsylvania State University, University Park, PA 16802

[7]Department of Physics, Applied Physics and Astronomy, Rensselaer Polytechnic Institute, Troy, NY 12180

[8]Department of Chemistry, The Pennsylvania State University, University Park, PA 16802

*Corresponding authors: mut11@psu.edu (M.T.); phanm@usf.edu (M.H.P.)

#These authors contributed equally to the work.

**Monolayers of molybdenum disulfide (MoS$_2$) are the most studied two-dimensional (2D) transition-metal dichalcogenides (TMDs), due to its exceptional optical, electronic, and opto-electronic properties. Recent studies have shown the possibility of incorporating a small amount of magnetic transition metals (e.g., Fe, Co, Mn, V) into MoS$_2$ to form a 2D dilute magnetic semiconductor (2D-DMS). However, the origin of the observed ferromagnetism has remained elusive, due to the presence of randomly generated sulfur vacancies during synthesis that can pair with magnetic dopants to form complex dopant-vacancy configurations altering the magnetic order induced by the dopants. By combining high-angle annular dark-field scanning transmission electron microscopy (HAADF-STEM) imaging with first-principles density functional theory (DFT) calculations and magnetometry data, we demonstrate the critical effects of sulfur vacancies and their pairings with vanadium atoms on the magnetic ordering in V-doped MoS$_2$ (V-MoS$_2$) monolayers. Additionally, we fabricated a series of field effect transistors on these V-MoS$_2$ monolayers and observed the emergence of p-type behavior as the vanadium concentration increased. Our study sheds light on the origin of ferromagnetism in V-MoS$_2$ monolayers and provides a foundation for future research on defect engineering to tune the electronic and magnetic properties of atomically thin TMD-based DMSs.**

Transition metal dichalcogenide (TMD) semiconducting monolayers $MX_2$ ($M$=Mo, W, etc.; $X$=S, Se, Te) display unique chemical, mechanical, and physical properties and are an important class of two-dimensional (2D) materials for a wide range of applications in field effect transistors (FETs), optoelectronics, spintronics, spin-caloritronics, and valleytronics.[1–6] However, the full potential of these 2D systems can only be realized through tuning their physical properties. Doping or alloying of transition metals into 2D-TMDs is an effective way to tailor their optical, electronic, and magnetic properties, thereby granting them new functionalities.[3,7,8] For example, vanadium doping/alloying in tungsten disulfide (V-WS$_2$), tungsten diselenide (V-WSe$_2$), and molybdenum diselenide (V-MoSe$_2$) monolayers behave as p-type semiconductors and exhibit room-temperature ferromagnetism.[9–13] This has resulted in a novel class of 2D diluted magnetic semiconductors (2D-DMSs). Interestingly, the magnetic properties of these 2D-DMSs can be electrically and optically manipulated at ambient temperature,[14–16] making them desirable for 2D vdW spintronics and opto-spintronics.[5,17,18] Alloying with concentrations of approximately 10 at.% have also been experimentally realized in WSe$_2$ and WS$_2$ with random V distributions.[9,10] However, the saturation magnetization ($M_S$) of these 2D-DMSs is limited to approximately $10^{-5}$ emu cm$^{-2}$, thus hampering their practical use. This has driven forward the search for new 2D-DMSs with enhanced magnetic functionalities.

Among many 2D-TMD candidates, molybdenum disulfide (MoS$_2$) has emerged as an excellent 2D semiconductor, due to its tunable charge-carrier types, high on/off ratio, high carrier mobilities, relatively large direct band gap, and strong spin-orbit coupling in the 2D limit.[1,19] This is why MoS$_2$ has been studied extensively and is considered one of the best alternatives to graphene, a prominent but gapless 2D material, for next-generation nanodevices.[19–22] Density functional theory (DFT) calculations performed by Fan *et.al.* for monolayer MoS$_2$ doped with the transition metals V, Mn, Fe, Co, and Cu have suggested that V doping/alloying (with concentration up to 8 at.%) is an effective way to induce and manipulate the magnetism in 2D-TMDs.[23] Gao *et. al.* have predicted an enhanced magnetization and high Curie temperature ($T_C$) in V-MoS$_2$ monolayers with high V concentrations (up to 9 at.%).[24] However, such a high alloying level is challenging to be achieved by chemical vapor deposition (CVD).[1,2] As compared to other TMD counterparts (e.g., WS$_2$, WSe$_2$),[9,10] the defective nature of CVD-grown MoS$_2$ monolayers also becomes a hindrance to exploring their magnetic properties.[25–30] In this context, Zhou *et. al.* carried out a systematic study on intrinsic structural defects including point defects, dislocations, grain boundaries, and edges, using atomic resolution HAADF-STEM imaging.[30] DFT calculations revealed that the unwanted structures can change the local density of states by introducing defect states in the bandgap and hence alter the materials' properties.. While the control of structural defects poses serious synthesis challenges, defective 2D-TMDs could potentially trigger lattice symmetry breaking and induce magnetic order. By intentionally introducing atomic sulfur vacancies (V$_S$) via Ar plasma irradiation, Hu *et. al.* observed a significant enhancement in the saturation magnetization of V-MoS$_2$ nanosheets when varying S-vacancy concentration.[27] The robust ferromagnetic response was attributed to the hybridization between the V 3d state and the V$_S$ impurity bands, according to the bound magnetic polaron model. However, the actual configurations of dopant-vacancy (V-S) pairs were not determined, and their effects on the magnetism in this 2D-TMD system were not understood. The ability to tune the electronic and magnetic functionalities of V-MoS$_2$ monolayers through different V-concentrations was also not achieved in previous studies,[27,31] although the V-concentration dependent optical properties were reported in the literature.[32] Room-temperature ferromagnetism was also reported in Fe-, Mn-, and Co-doped MoS$_2$ monolayers.[33–36] Interestingly, Huang *et al.* demonstrated the existence of spin transfer torque in a 2D Co-MoS$_2$/Ta heterostructure through spin Hall magnetoresistance (SMR) measurements.[34] However, the effects of chalcogen vacancies and dopant-vacancy pairs on the magnetic ordering in these 2D systems remain an open question.

To address these important, yet unresolved issues, a liquid-phase precursor-assisted CVD method[37] was used to achieve the controlled substitutional incorporation of vanadium atoms in monolayer MoS$_2$. The Raman modes activated by defects were identified, and magnetometry measurements at room temperature indicated an optimal ferromagnetic signal at vanadium concentrations of 2-3 at%. The combination of magnetometry measurements with HAADF-STEM imaging and first-principles DFT calculations, allowed for the investigation of the interaction between vanadium atoms and sulfur vacancies, and their contribution to the net magnetization. Furthermore, the fabrication of field effect transistors (FETs) on pristine and V-MoS$_2$ monolayers with varying V concentrations revealed the emergence of p-type behavior with increasing V concentration.

A scheme of the liquid-phase precursor-assisted CVD synthesis method is illustrated in Fig. S1a, and its details are given in the Methods section. The advantage of this method is that by tuning the V precursor solution concentration, the V concentration in MoS$_2$ can be controlled. As one can see in Fig. S1b, a low-magnification scanning electron microscope (SEM) image of the 3 at.% V-MoS$_2$ monolayer, shows that the coverage and distribution of flakes are uniform over large areas (few hundreds of μm$^2$), indicating a great scale-up potential of the liquid-phase precursor-assisted CVD doping/alloying method. Figure S1c shows a high-magnification SEM image of a single V-MoS$_2$ flake revealing the jagged morphology of flake edges. In Fig. S1d, the Raman spectra of V-MoS$_2$ monolayers shows the emergence of the following modes: TA(M), TA(K), LA(M), LA(K), and higher order modes. These modes were absent in pristine MoS$_2$, due to enhanced scattering from more defects (e.g., S vacancies, dopant-vacancy pairs) in the V incorporated samples. In Fig. S1e, the photoluminescence (PL) spectra of $MoS_2$ monolayers with different V concentrations shows a clear quenching of the PL intensity, in agreement with previous studies.[9,10,32] X-ray photoelectron spectra (XPS) were also collected (see Fig. S2). Here, we observe that the Mo 3d and S 2p peaks shift monotonically after increasing the V concentration, indicating the incorporation of V atoms into the $MoS_2$ lattice.

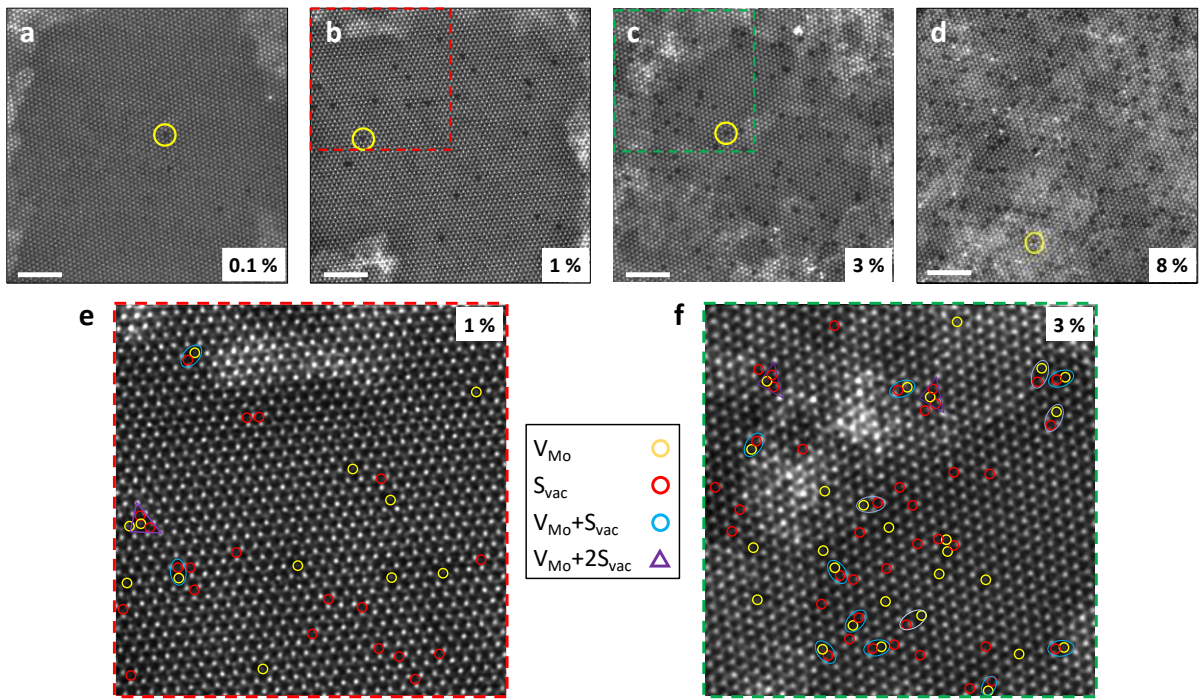

**Figure 1.** HAADF-STEM imaging of V-$MoS_2$ monolayers. **(a-d)** High-resolution STEM images of V atoms incorporated in $MoS_2$ with V concentration of 0.1, 1, 3, and 8 at%, respectively. Scale bar: 2 nm. Sulfur vacancy analysis upon each zoom-in region (red and green squares) for 1 at% **(e)** and 3 at% **(f)** V-$MoS_2$ monolayers reveal the abundance of S-vacancies ($S_{vac}$, red circles) forming near the substitutional V atoms ($V_{Mo}$, yellow circles). Beside the single- and double-

dopants, two major V-vacancy configurations were probed and represented by the blue circles ($V_{Mo}+S_{vac}$) and purple triangles ($V_{Mo}+2S_{vac}$).

The successful incorporation of V atoms into the MoS$_2$ lattice was confirmed via HAADF-STEM imaging to acquire atomic-resolution Z-contrast STEM images. As the V precursor solution concentration was increased, the V concentration was controllably increased from 0.1 at.% up to 8 at.%, as shown in Figs. 1a-d. To understand the presence of defect-activated Raman modes upon V incorporation, a systematic analysis of the defects present in the V-MoS$_2$ samples was performed on the HAADF-STEM images, as shown in Figs. 1e,f. Due to the Z-contrast of HAADF-STEM images, line scans of image intensities were investigated to carefully identify the sulfur vacancies ($S_{vac}$). As the V concentration increased, more sulfur vacancies were found in the lattice, which potentially explained the increased normalized intensities of the defect-activated Raman modes (TA(M), TA(K), LA(M), LA(K), and higher order modes). Furthermore, shown in the magnified HAADF-STEM images for 1 at.% and 3 at.% V in MoS$_2$ (Fig. 1e,f), we notice that $S_{vac}$ vacancies tend to be located next to the V atom sites and form dopant-vacancy clusters. The same HAADF-STEM analysis was also performed for the 2 at.% V-MoS$_2$ sample (Fig. S3). We observed two major configurations, which include a single V atom coupled with a single S-vacancy ($V_{Mo}$-$S_{vac}$, highlighted by the blue circles) and single V atom coupled with two nearest S-vacancies ($V_{Mo}$-$2S_{vac}$, highlighted by the purple triangles). Interestingly, when more V atoms were introduced to the MoS$_2$ lattice, not only did the V atoms get closer to each other, but the numbers of V-vacancy pairs/clusters also increased considerably, as seen in Figs. 1e,f. As shown below, the presence of these V-vacancy pairs also altered the electronic and magnetic properties of the V-MoS$_2$ monolayers when the V concentration was increased.

To probe the effects of V atoms, S vacancies, and V-vacancy pairs on the transport properties of $MoS_2$ monolayers, we fabricated a series of FETs on pristine $MoS_2$ and V-$MoS_2$ monolayers. Figures. 2a-c display the performances of these devices. It can be seen in Fig. 2a that the dual sweep transfer characteristics show little hysteresis, indicating the presence of very few interfacial trap states. As seen in Fig. 2b, the threshold voltages of the devices are shifted monotonically as the V concentration increases. At 3 at.% V concentration, the device turned to be ambipolar. However, the introduction of more scattering centers such as V atoms and sulfur vacancies inevitably reduced the device output and mobility (see Fig. 2c).[38]

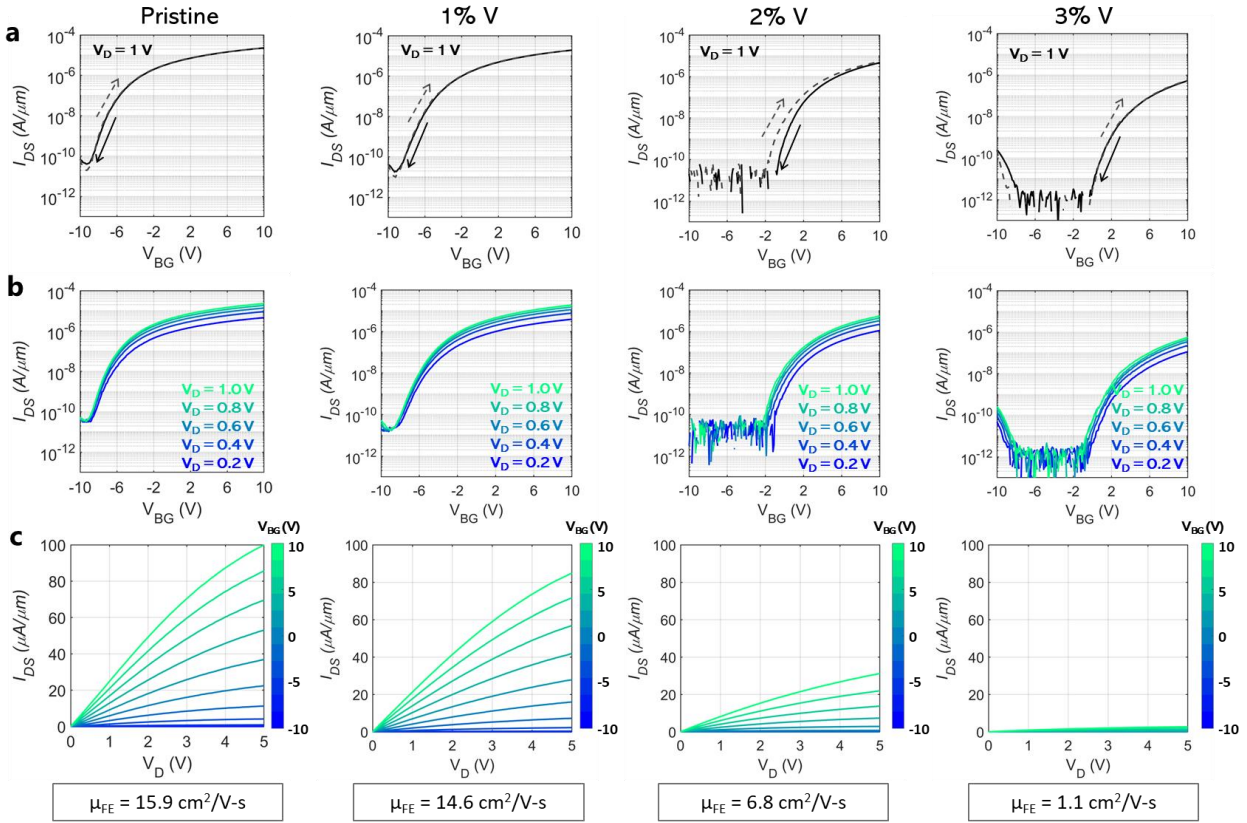

**Figure 2.** FET device performances for pristine $MoS_2$, 1 at.% V-$MoS_2$, 2 at.% V-$MoS_2$, and 3 at.% V-$MoS_2$, in terms of dual sweep transfer characteristics **(a)**, transfer characteristics **(b)**, and output

characteristics (**c**). Field-effect electron mobility (μ$_{FE}$) was also extracted from peak transconductance.

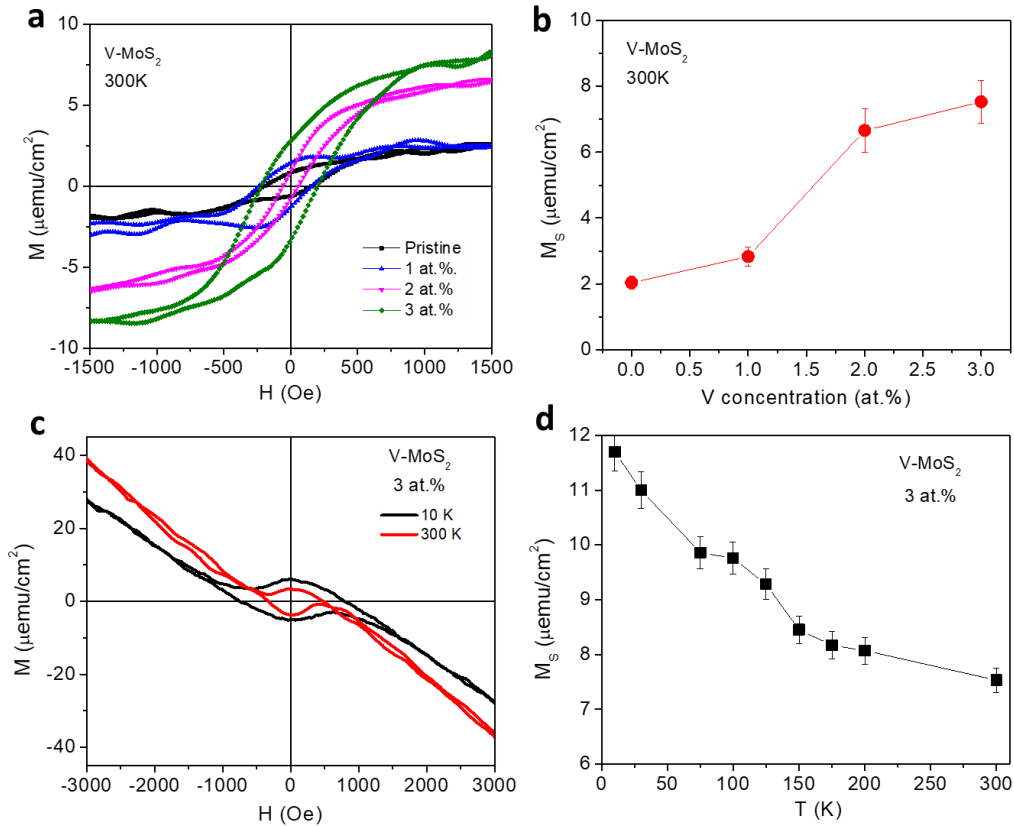

**Figure 3.** Magnetometry measurements on pristine and V-MoS$_2$ monolayers. (**a**) Magnetic hysteresis loops taken at 300 K for pristine and V-MoS$_2$ monolayers for 1 at.%, 2 at.%, and 3 at.% V concentrations. The *M-H* loops have had the diamagnetic background of the Si/SiO$_2$ substrate subtracted; (**b**) Saturation magnetization *M$_S$* of the V-MoS$_2$ monolayers is plotted as a function of V concentration. (**c**) A comparison of the *M-H* loops for the 3 at.% V-MoS$_2$ monolayer sample at 10 and 300 K; (**d**) Temperature dependence of *M$_S$* for the 3 at.% V-MoS$_2$ monolayer sample.

To probe the effects of V atoms, S vacancies, and V-vacancy (V-S) pairs on the magnetic properties of the V-MoS$_2$ monolayers, we performed systematic vibrating sample magnetometry (VSM) measurements and presented the results in Figs. 3a-d and Supporting Information (Figs. S4-6). As can be seen in Fig. 3a, at room temperature, pristine $MoS_2$ monolayers possess a weak magnetic ordering on their diamagnetic background. This has been previously attributed to the presence of defects in the form of S/Mo vacancies (e.g., line defects and/or a single vacancy Mo with two-sulfur vacancies $V_{Mo}$-$V_{2S}$) in $MoS_2$. [25,39–46] The incorporation of V atoms into the $MoS_2$ monolayers is found to enhance the ferromagnetism in $MoS_2$ monolayers at room temperature (Fig. 3a,b). The enhancement of ferromagnetism is more pronounced when more V atoms were introduced into the $MoS_2$ lattice up to 3 at.%, as demonstrated by a clear hysteresis loop (Fig. 3a) and a greater value of saturation magnetization $M_S$ (3-4 times larger) as compared to pristine $MoS_2$. Values of $M_S$ extracted from the $M(H)$ loops (Fig. 3a) are plotted as a function of V concentration (Fig. 3b).

It can be observed in Fig. 3b that when a small amount of V atoms (~1 at.%) was introduced into the $MoS_2$ system, we observed a slight increase in $M_S$. However, this modest change was significantly amplified as the V concentration doubled or tripled to 2 at.% and 3 at.%, respectively. Such a drastic enhancement in $M_S$ implies the establishment of the long-range FM order driven by V-V ferromagnetic (FM) interactions which become dominant in $MoS_2$ samples with higher V concentrations. However, very high V concentrations (e.g., 8 at.%) strongly suppresses the ferromagnetic ordering (Fig. S4). A similar trend has been reported for the V-$WSe_2$, V-$WS_2$, and V-$MoSe_2$ monolayers.[9,10,13] Our findings also demonstrate the intrinsic ferromagnetism in V-doped 2D-TMD semiconductors.

Among the samples investigated, the 3 at.% V-$MoS_2$ monolayer shows the largest values of $M_S \cong 7.5$ µemu/cm$^2$ and $H_C \cong 250$ Oe at room temperature (Fig. 3a). As expected, both $M_S$ and $H_C$ increase with lowering temperature for this sample (Figs. 3c,d). The absence of a shift in the field-cooled $M(H)$ loop with respect to the zero-field-cooled $M(H)$ loop also points to the dominance of V-V FM interactions in the V-MoS$_2$ monolayer (Fig. S5). Noticeably, the $M_S$ increases gradually with lowering temperature for $T >$ ~150 K and shows a sharp increase for $T <$ ~150 K. A similar feature has also been observed for the temperature-dependent magnetization $M(T)$ curves (Fig. S6). The temperature dependence of $M_S$ (Fig. 3d) can be explained by considering possible differences in strength of FM interactions between V atoms at closer and farther distances, as V atoms are randomly distributed throughout the V-MoS$_2$ monolayer (see Fig. 1f). It has been suggested that the ferromagnetism in V-doped TMD monolayers originates mainly from the Ruderman–Kittel Kasuya–Yosida (RKKY) interaction (determined by V-V distances).[9,10,23] At closer V-V distances, FM interactions are greater, driving the system to order magnetically at a higher temperature. In contrast, at farther V-V distances, FM interactions are weaker, causing the system to order magnetically at a lower temperature. Therefore, the remarkable change in $M_S$ around 150 K (Fig. 3d) can be attributed to enhanced FM interactions between V atoms at farther distances, in addition to the FM interactions between V atoms at closer distances. The unique combination of HAADF-STEM (Fig. 1f), VSM (Fig. 3d), and DFT (Fig. 4) results fully support this claim. Our findings may also shed light on the complexity of the temperature dependence of magnetization in other magnetically doped TMD monolayer systems.[9,10,13,27,47] Nonetheless, the magnetic properties of the V-MoS$_2$ monolayers cannot be explained solely by looking at FM interactions between V atoms, since a significant amount of defects in the form of S vacancies have been identified by HAADF-STEM imaging (Fig. 1e,f and Fig. S3).

The presence of S vacancies near V atoms can affect the magnetic ordering of V moments and V-V interactions in the V-MoS$_2$ monolayers. While it is challenging to solely reveal this effect from the VSM measurements (Fig. 3), we have performed first-principles calculations based on DFT to elucidate the various combinations of defects and V atoms, their proximity to each other, and their effects on the electronic and magnetic structures of monolayer MoS$_2$. We considered a series of systems based on an 8×8 MoS$_2$ supercell (64 Mo atoms and 128 S atoms). Multiple structures, such as a single substitutional V atom on the Mo site at the center of the supercell corresponding to the V concentration of ∼1.5 at.% (Fig. 4a, taken as reference) as well as two V atoms with varying distances between them representing the V concentration of ∼3 at.% (Figs. 4b-e), are simulated. Several structures with one V substitutional atom paired with one vacancy on the S atom site (Figs. 4f-i) as well as one V atom paired with two S vacancies (Figs. 4j-m) are also calculated.

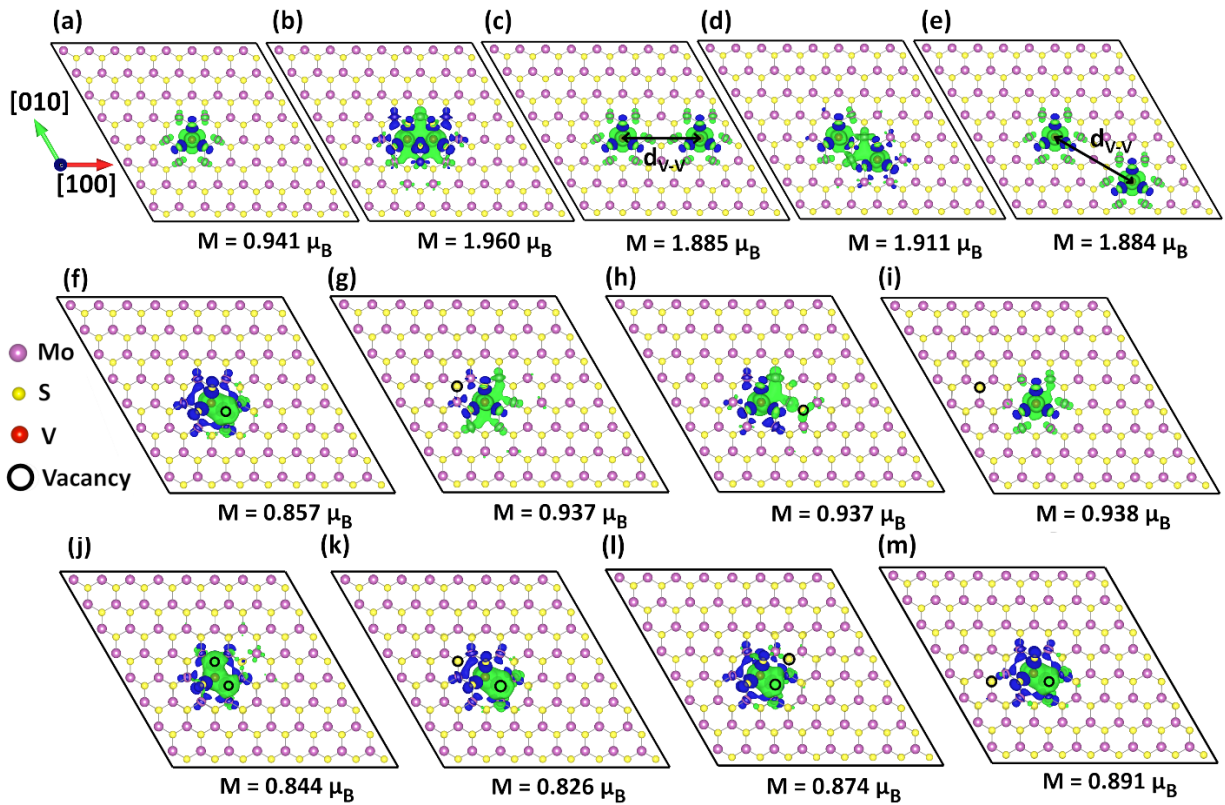

**Figure 4.** Spin density of MoS$_2$ monolayer modeled by a supercell with 192 atoms: **(a)** a V atom substituting a Mo atom, **(b-e)** two V atoms substituting two Mo atoms separated by different

distances dv-v; **(f-i)** a V atom substituting a Mo atom (under the green region) and a vacancy on the S site; **(j-m)** a V atom substituting a Mo atom and two vacancies on the S sites. Green and blue isosurfaces represent spin-up and spin-down densities (isosurface value at 0.001 e/Å$^3$).

The distance between the V atoms in the supercell is varied as illustrated in Figs 4.b-e; the V atoms can substitute adjacent or distant Mo atoms along the (100) or (1$\bar{1}$0) directions tracked by the distance separation $d_{V-V}$. We find that in all cases, the V-MoS$_2$ monolayer is a ferromagnet, which agrees with our experimental observation (Fig. 3) and with that of previous studies.[23] The local spin distribution, as obtained from the simulations, helps us understand how magnetic ordering occurs. In the case of a single V in the supercell (Fig. 4a), we find that the nearest V-Mo spins couple ferromagnetically, while the V-S spins are arranged antiferromagnetically. This feature is similar to that observed for the V-WS$_2$ monolayers[9] but different from that observed for the V-WSe$_2$ monolayers in which the nearest V-W spins couple antiferromagnetically.[10] Increasing the number of V atoms leads to stronger ferromagnetism in the V-MoS$_2$ monolayer (Figs. 4b-e); the total magnetization increases from $M = 0.941\ \mu_B$ for a single V atom (Fig. 4a) to $M = 1.960\ \mu_B$ for two adjacent V atoms (Fig. 4b). It is found that the V-V separation also affects the total magnetization $M$. For adjacent V atoms located along (100) $M = 1.960\ \mu_B$, while when the separation of the adjacent V atoms is along (1$\bar{1}$0), $M = 1.911\ \mu_B$. As $d_{V-V}$ increases, the FM coupling between the two V atoms diminishes (Fig. 4b,c and Fig. 4d,e). It is also interesting to compare the structural stability of the different systems simulated here. We find that the structures with adjacent V atoms and S vacancies have lower total DFT energy $E_{tot}$ compared to those in which there is a longer separation between them. For example, $E_{tot}$ corresponding to the structure from Fig. 4f is lower by 251.7 meV compared to the one for Fig. 4j. Also, $E_{tot}$ corresponding to the structure from Fig. 4j is lower by 97.2 meV compared to the one for Fig. 4l. The energy-favored structural stability of dopants and vacancies being on the adjacent sites may indicate that such configurations are easier to form experimentally (Fig. 1f), thus they occur more frequently in the synthesis process.

The results for the directional and separational dependence of the magnetization are summarized in the upper panel of Fig. 5a, where we distinguish between the total $M$ and the magnetization associated with the V atoms $M_V$. The dashed lines represent the doubled values of the magnetization of the structure with a single V atom (about 1.5 at.% V concentration) and serve as reference lines for comparison with the magnetizations of the structures with two V atoms (about 3 at.% V concentration) in different directions and separation distances . We note that typically $M > M_V$ except when the two V atoms are their nearest neighbors. Each isolated V atom and its nearest surrounding Mo, S atoms creates a magnetic domain of radius around 3.17 Å (see Fig. 5a). When $d_{V-V}$ is twice as large as this domain size, there is almost no overlap between the magnetic domains of two V atoms. The isolation of the domain regions in this case is reflected in the asymptotic behavior of $M$ and $M_V$ of structures with 3 at.% V concentration which almost equal to twice of the $M$ and $M_V$ of the structure with 1.5 at.% V concentration when the distance between two V atoms is larger than 7 angstroms.

Simulations of defective and V-MoS$_2$ monolayers are also performed. In the supercell from Fig. 4f-i, we consider a V atom and a vacancy on the S site. It is important to note that the magnetic ordering in MoS$_2$ cannot be driven solely by single S vacancies. This is consistent with previous studies,[25–27,48] showing that one needs to create vacant disturbances on the Mo atom to induce magnetic states. It appears, however, that the interplay between V–S vacancies and their proximity affects the local spin distribution and overall magnetization. When the S vacancy and V atom are next to each other, the local spin distribution becomes notably disturbed when compared with the alternating local FM-AFM coupling of a single vanadium atom (Fig. 4f). The local magnetic moment of V for this case increases to 1.013 $\mu_B$ and the nearest Mo-site and S-site spins couple antiferromagnetically with the V site spin, except for the S site in the opposite face with the S vacancy. This trend is different from that found for the case of V-WSe$_2$ monolayers, in which the presence of mono S-vacancies does not affect the local moment of a V atom and the V-V FM coupling.[12,49] Insertion of one more S vacancy into the case of one S vacancy near a V atom insignificantly changes the magnetic configuration and the net magnetic moment of the V-MoS$_2$ system, thus indicating that the dramatic role of S vacancies is only relevant at nearest neighbor length scales (Figs. 4j-m). These findings not only explain the underlying origin of the magnetism in the presently studied V-MoS$_2$ samples (Fig. 3), but also shed light on the complexity of the S-vacancy concentration dependence of magnetization in the V-MoS$_2$ nanosheets reported by Hu *et al.*[27]

The V atoms and S vacancies affect not only the local spin distributions, but their signatures are also found in the electronic properties. When 1.5 at% V is introduced into the MoS$_2$ supercell, there is one spin-down defect state near the maximum valence band and one spin-up defect state near the minimum conduction band, but the V-MoS$_2$ monolayer remains semiconducting (Fig. 5b). On the other hand, doubling the V concentration transforms the monolayer into a half-metallic state. The splitting of the defect states due to the proximity of V atoms is accompanied by shifting of one of the two spin-down defect states closer to the valence band maximum making it partially filled (Figs. 5c,d). This can explain the ambipolar characteristics of the FET device made of the 3 at.% V-MoS$_2$ monolayer (Fig. 2c). Moving two V atoms away from each other not only reduces the complexity of the moment distribution on the sample but also reduces its half-metallicity.

Indeed, as can be seen in the bottom panel of Fig. 5a, density of states at Fermi level for spin-down component dramatically reduces when increasing V-V separation.

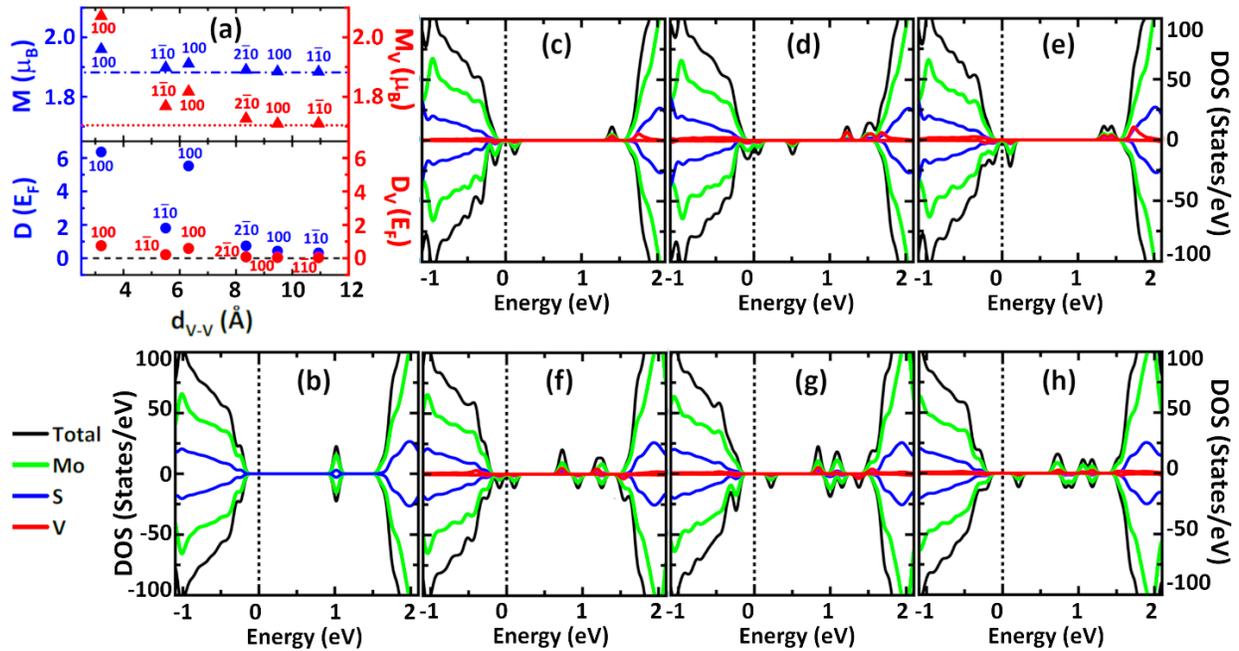

**Figure 5.** Density of states (DOS) of different V configurations in V-MoS$_2$ monolayers **(a)** Total magnetization $M$ ($\mu_B$) and magnetic moments located on V atoms $M_V$ ($\mu_B$) (upper panel); total density of states at Fermi level $D(E_F)$ and density of states due to V atoms at Fermi level $D_V(E_F)$ (bottom panel); dashed lines represent the doubled magnetization for the structure with a single V atom (about 1.5 at.% V concentration) and serve as reference lines for comparison. Total and atomically resolved density of states for **(b)** single V atom in MoS$_2$ monolayer supercell as in Figure 5a; **(c,d)** two V atoms in MoS$_2$ monolayer supercell with different positions as in Figs. 4**b,c**, respectively; **(e)** one S vacancy in MoS$_2$ monolayer supercell; **(f-h)** one V atom with two S vacancies in MoS$_2$ monolayer supercell with different positions as in Figs. 4(**j-l**), respectively.

The removal of an S atom in the monolayer supercell also introduces impurity states in the semiconducting region of the monolayer as shown in Fig. 5e. However, unlike the case of V atoms, the spin-up, and spin-down channels, arising from the 4d orbitals of the Mo atoms, are completely symmetrical. Introducing V atoms into the defective MoS$_2$ monolayer causes the occurrence of several impurity states in the band gap region. For close dopant-vacancy proximity, the stronger V-S-Mo interaction causes the spread of spin-up and down DOS peaks in the vicinity of the conduction band edge. Near the valence band edge, spin-down peaks are found in DOS originating from the coupling between V atoms and its nearest neighbor S vacancies, but no such feature is obtained for spin-up DOS (Fig. 5f-h).

In summary, HAADF-STEM measurements indicate an increasing population of V-V and V-S$_{vac}$ pairs at higher V-concentrations, which supports the monotonic increase in the saturation magnetization of V-MoS$_2$ monolayers as observed via VSM measurements, up to 3 at.%. Unlike V-WS$_2$ and V-WSe$_2$ systems, in which the magnetic moment vanishes at the nearest V-V pairing, the opposite behavior is observed in V-MoS$_2$, which offers the possibility of tuning the magnetic properties in 2D-TMDs by combining long-range and short-range magnetic ordering. In defective MoS$_2$ monolayers, the universal configuration of single dopant and adjacent chalcogen vacancies plays an important role in the overall magnetism and should be considered carefully and equivalently to other major dopant complexes. Our findings shed light on the magnetic complexity of MoS$_2$ monolayer when doped/alloyed with magnetic transition metals (V, Fe, Co, Mn)[26–29,33–36] and pave a pathway for tuning the electronic and magnetic properties of 2D-TMD based DMSs through defect engineering.

## Methods

*Synthesis*

Solution 1 was prepared by dissolving ammonium molybdate tetrahydrate (($NH_4$)$_6Mo_7O_{24}$ · $4H_2O$, Sigma-Aldrich, purity of 99.98%; 0.01g) and sodium cholate hydrate ($C_{24}H_{39}NaO_5$ · $xH_2O$, Sigma-Aldrich, purity of ≥99%; 0.2 g) into DI water (10 mL). Solution 2 was prepared by dissolving vanadium (IV) oxide sulfate hydrate ($VOSO_4 \cdot xH_2O$, Sigma-Aldrich, purity of ≥99.99%, up to 0.02g) into DI water (10 mL). Freshly mixed Sol 1 (1 mL) and Sol 2 (1 mL) were spin-coated onto a freshly cleaned and plasma-treated $SiO_2$ (300 nm)/Si substrate (3000 rpm, 1 min). Subsequently, the precursor-coated substrate was loaded in the center of a quartz reaction tube, and 300-400 mg of sulfur powders was placed in an alumina boat located upstream. During synthesis, the furnace was ramped up to 800 °C in 30 min and held for 15 min. Simultaneously, the sulfur powders were vaporized at 220 °C, allowing the sulfurization of the spin-coated precursors. Ar was used as the carrier gas throughout the process, with a flow rate of 100 sccm. The synthesis of pristine $MoS_2$ is identical to that of V-$MoS_2$, except that only Solution 1 is spin-coated onto the substrate.

*Characterizations*

Scanning electron microscopy images were taken with a field emission gun-equipped Apreo instrument (Thermo Scientific) with an ETD detector at 5kV and 50pA current.

The Raman and PL spectrums were acquired using a Horiba LabRAM HR Evolution Raman Microscope with a 532nm excitation laser in a backscattering configuration and 1800⁻ or 600⁻ line/mm gratings.

XPS experiments were performed using a Physical Electronics VersaProbe II instrument equipped with a monochromatic Al kα x-ray source (hν = 1,486.7 eV) and a concentric hemispherical analyzer. The binding energy axis was calibrated using sputter cleaned Cu (Cu 2p3/2 = 932.62 eV, Cu 3p3/2 = 75.1 eV) and Au foils (Au 4f7/2 = 83.96 eV). Measurements were made on the as-grown samples, at a takeoff angle of 45° with respect to the sample surface plane and using a beam size of 200 x 200 µm. Charge neutralization was performed using both low energy electrons (<5 eV) and argon ions. XPS spectra was fitted using CasaXPS v2.3.25PR1.0. Peaks on all samples/regions were charge-referenced to adventitious carbon at 284.8 eV

The V-MoS$_2$ samples were transferred from the growth substrate (Si/SiO$_2$) onto Quantifoil R 2/1 200 gold mesh TEM grids. High-angle annular dark-field (HAADF)-STEM was then performed on an FEI Titan G2 60–300 kV microscope with a cold field-emission electron source. The microscope was operating at 80 kV with double spherical aberration correction to obtain a high resolution of ≈0.7 Å. Z-contrast STEM images were acquired using a HAADF detector with a collection angle of 42-244 mrad, camera length of 115 mm, beam current of 46 pA, and beam convergence angle of 30 mrad.

*Electronic device fabrication and measurements*

Pristine MoS$_2$ and V-MoS$_2$ flakes were transferred from the growth substrate (Si/SiO$_2$) to a 50 nm thick and atomic layer deposition grown Al$_2$O$_3$ substrate with Pt/TiN/p++Si as the back-gate electrode. All FETs were fabricated with a channel length of 1 µm with 40 nm Ni/30 nm Au as the source/drain contact electrodes defined using a standard electron-beam lithography process.

*Magnetic measurements*

Temperature- and magnetic-field-dependent magnetization measurements were carried out in a Physical Property Measurement System (PPMS) from Quantum Design with a vibrating sample magnetometer (VSM) probe. For magnetic field-dependent magnetization measurements *M(H)*, the sample was cooled from 300 K in the absence of a dc field and *M*(*H*) data was taken at each selected measurement temperature. To examine if FM interactions between V atoms are dominant in the V-MoS$_2$ monolayers, *M*(*H*) loops were collected at 10 K under zero-field-cooled (ZFC) and field-cooled (FC) measurement protocols by cooling down the sample from 300 K to 10 K in the absence and presence of a magnetic field of 0.5 T, respectively. Temperature-dependent magnetization measurements *M*(*T*) were also carried out over 10-350 K under ZFC and FC protocols in a field of 0.5 T.

*Computational Methods*

All structures are investigated by DFT calculations. The generalized gradient approximation (GGA) with Perdew–Burke–Ernzerhof (PBE) for the exchange-correlation energy is used, as implemented in the Vienna ab initio simulations package (VASP).[50–52] We construct an $8 \times 8 \times 1$ MoS$_2$ supercell from a primitive cell representation (64 Mo and 128 S atoms). The electronic shell structure is taken to have 6 valence electrons of Mo (4d$^5$ 5s$^1$), 6 valence electrons of S (3s$^2$ 3p$^4$), and 5 valence electrons of V (3d$^3$ 4s$^2$). The kinetic energy cutoff is 450 eV. To consider vdW interactions, the DFT-D3(BJ) method also has been integrated with this study.[53] The relaxation criteria of all systems are used by allowing all cell parameters to change with 10$^{-5}$ eV energy and 0.01 eV/Å force relaxation criteria. The integration over the Brillouin zone is performed at $2 \times 2 \times 1$ and $9 \times 9 \times 1$ Γ-centered Monkhorst-Pack k-point meshes for optimizations and density of states, respectively. These calculations are performed by employing Gaussian smearing with 0.05 eV width. The GGA+U method within Dudarev's approach is employed to account for the strongly correlated electronic systems.[54] The Hubbard potential U value chosen for the d-orbital of V is 3.0 eV.[14,48] A vacuum of 15 Å is considered on each slab to avoid the interaction between periodic images. The VESTA package is employed to visualize the model structures.[55] VASPKIT code is used for postprocessing calculated data.[56]

## Acknowledgements

Work at PSU was supported by the Air Force Office of Scientific Research (AFOSR) through grant No. FA9550-23-1-0447 and the NSF-IUCRC Center for Atomically Thin Multifunctional Coatings (ATOMIC). M.H.P acknowledges support from the US Department of Energy, Office of Basic Energy Sciences, Division of Materials Sciences and Engineering under Award No. DE-FG02-07ER46438. D.T.-X.D. would like to acknowledge support from the Presidential Fellowship sponsored by the University of South Florida. L.M.W. acknowledges support from the US Department of Energy under Grant No. DE-FG02-06ER46297. Computational resources are provided by USF Research Computing. Noah Schulz is acknowledged for proofreading the manuscript.

## Conflict of Interest

The authors declare no conflict of interest.

# References

(1) Choi, W.; Choudhary, N.; Han, G. H.; Park, J.; Akinwande, D.; Lee, Y. H. Recent Development of Two-Dimensional Transition Metal Dichalcogenides and Their Applications. *Mater. Today* **2017**, *20* (3), 116–130.

(2) Lei, Y.; Zhang, T.; Lin, Y.-C.; Granzier-Nakajima, T.; Bepete, G.; Kowalczyk, D. A.; Lin, Z.; Zhou, D.; Schranghamer, T. F.; Dodda, A.; Sebastian, A.; Chen, Y.; Liu, Y.; Pourtois, G.; Kempa, T. J.; Schuler, B.; Edmonds, M. T.; Quek, S. Y.; Wurstbauer, U.; Wu, S. M.; Glavin, N. R.; Das, S.; Dash, S. P.; Redwing, J. M.; Robinson, J. A.; Terrones, M. Graphene and Beyond: Recent Advances in Two-Dimensional Materials Synthesis, Properties, and Devices. *ACS Nanosci. Au* **2022**, *2* (6), 450–485.

(3) Lin, Y.-C.; Torsi, R.; Younas, R.; Hinkle, C. L.; Rigosi, A. F.; Hill, H. M.; Zhang, K.; Huang, S.; Shuck, C. E.; Chen, C.; Lin, Y.-H.; Maldonado-Lopez, D.; Mendoza-Cortes, J. L.; Ferrier, J.; Kar, S.; Nayir, N.; Rajabpour, S.; Van Duin, A. C. T.; Liu, X.; Jariwala, D.; Jiang, J.; Shi, J.; Mortelmans, W.; Jaramillo, R.; Lopes, J. M. J.; Engel-Herbert, R.; Trofe, A.; Ignatova, T.; Lee, S. H.; Mao, Z.; Damian, L.; Wang, Y.; Steves, M. A.; Knappenberger, K. L.; Wang, Z.; Law, S.; Bepete, G.; Zhou, D.; Lin, J.-X.; Scheurer, M. S.; Li, J.; Wang, P.; Yu, G.; Wu, S.; Akinwande, D.; Redwing, J. M.; Terrones, M.; Robinson, J. A. Recent Advances in 2D Material Theory, Synthesis, Properties, and Applications. *ACS Nano* **2023**, *17* (11), 9694–9747.

(4) Sierra, J. F.; Fabian, J.; Kawakami, R. K.; Roche, S.; Valenzuela, S. O. Van Der Waals Heterostructures for Spintronics and Opto-Spintronics. *Nat. Nanotechnol.* **2021**, *16* (8), 856–868.

(5) Phan, M.-H.; Trinh, M. T.; Eggers, T.; Kalappattil, V.; Uchida, K.; Woods, L. M.; Terrones, M. A Perspective on Two-Dimensional van Der Waals Opto-Spin-Caloritronics. *Appl. Phys. Lett.* **2021**, *119* (25), 250501.

(6) Mak, K. F.; Xiao, D.; Shan, J. Light–Valley Interactions in 2D Semiconductors. *Nat. Photonics* **2018**, *12* (8), 451–460.

(7) Cai, L.; Tung, V.; Wee, A. Room-Temperature Ferromagnetism in Two-Dimensional Transition Metal Chalcogenides: Strategies and Origin. *J. Alloys Compd.* **2022**, *913*, 165289.

(8) Xiong, Y.; Xu, D.; Feng, Y.; Zhang, G.; Lin, P.; Chen, X. P-Type 2D Semiconductors for Future Electronics. *Adv. Mater.* **2023**, 2206939.

(9) Zhang, F.; Zheng, B.; Sebastian, A.; Olson, D. H.; Liu, M.; Fujisawa, K.; Pham, Y. T. H.; Jimenez, V. O.; Kalappattil, V.; Miao, L.; Zhang, T.; Pendurthi, R.; Lei, Y.; Elías, A. L.; Wang, Y.; Alem, N.; Hopkins, P. E.; Das, S.; Crespi, V. H.; Phan, M.; Terrones, M. Monolayer Vanadium-Doped Tungsten Disulfide: A Room-Temperature Dilute Magnetic Semiconductor. *Adv. Sci.* **2020**, *7* (24), 2001174.

(10) Pham, Y. T. H.; Liu, M.; Jimenez, V. O.; Yu, Z.; Kalappattil, V.; Zhang, F.; Wang, K.; Williams, T.; Terrones, M.; Phan, M. Tunable Ferromagnetism and Thermally Induced Spin Flip in Vanadium-Doped Tungsten Diselenide Monolayers at Room Temperature. *Adv. Mater.* **2020**, *32* (45), 2003607.

(11) Yun, S. J.; Duong, D. L.; Ha, D. M.; Singh, K.; Phan, T. L.; Choi, W.; Kim, Y.; Lee, Y. H. Ferromagnetic Order at Room Temperature in Monolayer $WSe_2$ Semiconductor via Vanadium Dopant. *Adv. Sci.* **2020**, *7* (9), 1903076.

(12) Yun, S. J.; Cho, B. W.; Dinesh, T.; Yang, D. H.; Kim, Y. I.; Jin, J. W.; Yang, S.; Nguyen, T. D.; Kim, Y.; Kim, K. K.; Duong, D. L.; Kim, S.; Lee, Y. H. Escalating Ferromagnetic Order via Se-Vacancies Near Vanadium in WSe$_2$ Monolayers. *Adv. Mater.* **2022**, *34* (10), 2106551.

(13) Deng, J.; Zhou, Z.; Chen, J.; Cheng, Z.; Liu, J.; Wang, Z. Vanadium-Doped Molybdenum Diselenide Atomic Layers with Room-Temperature Ferromagnetism. *ChemPhysChem* **2022**, *23* (16), e202200162.

(14) Duong, D. L.; Yun, S. J.; Kim, Y.; Kim, S.-G.; Lee, Y. H. Long-Range Ferromagnetic Ordering in Vanadium-Doped WSe2 Semiconductor. *Appl. Phys. Lett.* **2019**, *115* (24), 242406.

(15) Ortiz Jimenez, V.; Pham, Y. T. H.; Liu, M.; Zhang, F.; Yu, Z.; Kalappattil, V.; Muchharla, B.; Eggers, T.; Duong, D. L.; Terrones, M.; Phan, M. Light-Controlled Room Temperature Ferromagnetism in Vanadium-Doped Tungsten Disulfide Semiconducting Monolayers. *Adv. Electron. Mater.* **2021**, *7* (8), 2100030.

(16) Jimenez, V. O.; Pham, Y. T. H.; Zhou, D.; Liu, M. Z.; Nugera, F. A.; Kalappattil, V.; Eggers, T.; Hoang, K.; Duong, D. L.; Terrones, M.; Gutierrez, H. R.; Phan, M. H. Transition Metal Dichalcogenides: Making Atomic-Level Magnetism Tunable with Light at Room Temperature. *Adv. Sci.* **2023**, 2304792.

(17) Nguyen, L.-A. T.; Jiang, J.; Nguyen, T. D.; Kim, P.; Joo, M.-K.; Duong, D. L.; Lee, Y. H. Electrically Tunable Magnetic Fluctuations in Multilayered Vanadium-Doped Tungsten Diselenide. *Nat. Electron.* **2023**, *6* (8), 582–589.

(18) Phan, M.-H. New Research Trends in Electrically Tunable 2D van Der Waals Magnetic Materials. **2023**.

(19) Gupta, D.; Chauhan, V.; Kumar, R. A Comprehensive Review on Synthesis and Applications of Molybdenum Disulfide ($MoS_2$) Material: Past and Recent Developments. *Inorg. Chem. Commun.* **2020**, *121*, 108200.

(20) Nalwa, H. S. A Review of Molybdenum Disulfide ($MoS_2$) Based Photodetectors: From Ultra-Broadband, Self-Powered to Flexible Devices. *RSC Adv.* **2020**, *10* (51), 30529–30602.

(21) Kumar, V. P.; Panda, D. K. Review—Next Generation 2D Material Molybdenum Disulfide ($MoS_2$): Properties, Applications and Challenges. *ECS J. Solid State Sci. Technol.* **2022**, *11* (3), 033012.

(22) Ren, H.; Xiang, G. Strain-Modulated Magnetism in MoS2. *Nanomaterials* **2022**, *12* (11), 1929.

(23) Fan, X.-L.; An, Y.-R.; Guo, W.-J. Ferromagnetism in Transitional Metal-Doped MoS2 Monolayer. *Nanoscale Res. Lett.* **2016**, *11* (1), 154.

(24) Gao, Y.; Ganguli, N.; Kelly, P. J. DFT Study of Itinerant Ferromagnetism in p-Doped Monolayers of $MoS_2$. *Phys. Rev. B* **2019**, *100* (23), 235440.

(25) Cai, L.; He, J.; Liu, Q.; Yao, T.; Chen, L.; Yan, W.; Hu, F.; Jiang, Y.; Zhao, Y.; Hu, T.; Sun, Z.; Wei, S. Vacancy-Induced Ferromagnetism of $MoS_2$ Nanosheets. *J. Am. Chem. Soc.* **2015**, *137* (7), 2622–2627.

(26) Wang, Y.; Tseng, L.-T.; Murmu, P. P.; Bao, N.; Kennedy, J.; Ionesc, M.; Ding, J.; Suzuki, K.; Li, S.; Yi, J. Defects Engineering Induced Room Temperature Ferromagnetism in Transition Metal Doped MoS 2. *Mater. Des.* **2017**, *121*, 77–84.

(27) Hu, W.; Tan, H.; Duan, H.; Li, G.; Li, N.; Ji, Q.; Lu, Y.; Wang, Y.; Sun, Z.; Hu, F.; Wang, C.; Yan, W. Synergetic Effect of Substitutional Dopants and Sulfur Vacancy in Modulating the Ferromagnetism of MoS$_2$ Nanosheets. *ACS Appl. Mater. Interfaces* **2019**, *11* (34), 31155–31161.

(28) Wang, J.; Sun, F.; Yang, S.; Li, Y.; Zhao, C.; Xu, M.; Zhang, Y.; Zeng, H. Robust Ferromagnetism in Mn-Doped MoS2 Nanostructures. *Appl. Phys. Lett.* **2016**, *109* (9), 092401.

(29) Xiang, Z.; Zhang, Z.; Xu, X.; Zhang, Q.; Wang, Q.; Yuan, C. Room-Temperature Ferromagnetism in Co Doped MoS $_2$ Sheets. *Phys. Chem. Chem. Phys.* **2015**, *17* (24), 15822–15828.

(30) Zhou, W.; Zou, X.; Najmaei, S.; Liu, Z.; Shi, Y.; Kong, J.; Lou, J.; Ajayan, P. M.; Yakobson, B. I.; Idrobo, J.-C. Intrinsic Structural Defects in Monolayer Molybdenum Disulfide. *Nano Lett.* **2013**, *13* (6), 2615–2622.

(31) Sahoo, K. R.; Panda, J. J.; Bawari, S.; Sharma, R.; Maity, D.; Lal, A.; Arenal, R.; Rajalaksmi, G.; Narayanan, T. N. Enhanced Room-Temperature Spin-Valley Coupling in V-Doped MoS$_2$. *Phys. Rev. Mater.* **2022**, *6* (8), 085202.

(32) Zhang, J.; Zhu, Y.; Tebyetekerwa, M.; Li, D.; Liu, D.; Lei, W.; Wang, L.; Zhang, Y.; Lu, Y. Vanadium-Doped Monolayer MoS$_2$ with Tunable Optical Properties for Field-Effect Transistors. *ACS Appl. Nano Mater.* **2021**, *4* (1), 769–777.

(33) Fu, S.; Kang, K.; Shayan, K.; Yoshimura, A.; Dadras, S.; Wang, X.; Zhang, L.; Chen, S.; Liu, N.; Jindal, A.; Li, X.; Pasupathy, A. N.; Vamivakas, A. N.; Meunier, V.; Strauf, S.; Yang, E.-H. Enabling Room Temperature Ferromagnetism in Monolayer MoS$_2$ via in Situ Iron-Doping. *Nat. Commun.* **2020**, *11* (1), 2034.

(34) Huang, M.; Xiang, J.; Feng, C.; Huang, H.; Liu, P.; Wu, Y.; N'Diaye, A. T.; Chen, G.; Liang, J.; Yang, H.; Liang, J.; Cui, X.; Zhang, J.; Lu, Y.; Liu, K.; Hou, D.; Liu, L.; Xiang, B. Direct

Evidence of Spin Transfer Torque on Two-Dimensional Cobalt-Doped MoS$_2$ Ferromagnetic Material. *ACS Appl. Electron. Mater.* **2020**, *2* (6), 1497–1504.

(35) Singh, A.; Price, C. C.; Shenoy, V. B. Magnetic Order, Electrical Doping, and Charge-State Coupling at Amphoteric Defect Sites in Mn-Doped 2D Semiconductors. *ACS Nano* **2022**, *16* (6), 9452–9460.

(36) Duan, H.; Guo, P.; Wang, C.; Tan, H.; Hu, W.; Yan, W.; Ma, C.; Cai, L.; Song, L.; Zhang, W.; Sun, Z.; Wang, L.; Zhao, W.; Yin, Y.; Li, X.; Wei, S. Beating the Exclusion Rule against the Coexistence of Robust Luminescence and Ferromagnetism in Chalcogenide Monolayers. *Nat. Commun.* **2019**, *10* (1), 1584.

(37) Zhang, T.; Fujisawa, K.; Zhang, F.; Liu, M.; Lucking, M. C.; Gontijo, R. N.; Lei, Y.; Liu, H.; Crust, K.; Granzier-Nakajima, T.; Terrones, H.; Elías, A. L.; Terrones, M. Universal *In Situ* Substitutional Doping of Transition Metal Dichalcogenides by Liquid-Phase Precursor-Assisted Synthesis. *ACS Nano* **2020**, *14* (4), 4326–4335.

(38) Zou, J.; Cai, Z.; Lai, Y.; Tan, J.; Zhang, R.; Feng, S.; Wang, G.; Lin, J.; Liu, B.; Cheng, H.-M. Doping Concentration Modulation in Vanadium-Doped Monolayer Molybdenum Disulfide for Synaptic Transistors. *ACS Nano* **2021**, *15* (4), 7340–7347.

(39) Li, A.; Pan, J.; Yang, Z.; Zhou, L.; Xiong, X.; Ouyang, F. Charge and Strain Induced Magnetism in Monolayer MoS2 with S Vacancy. *J. Magn. Magn. Mater.* **2018**, *451*, 520–525.

(40) Yun, W. S.; Lee, J. D. Strain-Induced Magnetism in Single-Layer MoS$_2$ : Origin and Manipulation. *J. Phys. Chem. C* **2015**, *119* (5), 2822–2827.

(41) Tao, P.; Guo, H.; Yang, T.; Zhang, Z. Strain-Induced Magnetism in MoS$_2$ Monolayer with Defects. *J. Appl. Phys.* **2014**, *115* (5), 054305.

(42) Xue, L.; He, C.; Yang, Z.; Zhang, Z.; Xu, L.; Fan, X.; Zhang, L.; Yang, L. Strain Induced Magnetic Hysteresis in $MoS_2$ and $WS_2$ Monolayers with Symmetric Double Sulfur Vacancy Defects. *Phys. Chem. Chem. Phys.* **2022**, *24* (28), 17263–17270.

(43) Li, H.; Huang, M.; Cao, G. Magnetic Properties of Atomic 3d Transition-Metal Chains on S-Vacancy-Line Templates of Monolayer $MoS_2$ : Effects of Substrate and Strain. *J. Mater. Chem. C* **2017**, *5* (18), 4557–4564.

(44) Kumar, R.; Kelkar, A. H.; Singhal, R.; Sathe, V. G.; Choudhary, R. J.; Shukla, N. Strain Induced Structural Changes and Magnetic Ordering in Thin $MoS_2$ Flakes as a Consequence of 1.5 MeV Proton Ion Irradiation. *J. Alloys Compd.* **2023**, *951*, 169882.

(45) Zhang, K.; Pan, Y.; Wang, L.; Mei, W.-N.; Wu, X. Extended 1D Defect Induced Magnetism in 2D $MoS_2$ Crystal. *J. Phys. Condens. Matter* **2020**, *32* (21), 215302.

(46) Sanikop, R.; Sudakar, C. Tailoring Magnetically Active Defect Sites in $MoS_2$ Nanosheets for Spintronics Applications. *ACS Appl. Nano Mater.* **2020**, *3* (1), 576–587.

(47) Shen, D.; Zhao, B.; Zhang, Z.; Zhang, H.; Yang, X.; Huang, Z.; Li, B.; Song, R.; Jin, Y.; Wu, R.; Li, B.; Li, J.; Duan, X. Synthesis of Group VIII Magnetic Transition-Metal-Doped Monolayer $MoSe_2$. *ACS Nano* **2022**, *16* (7), 10623–10631.

(48) Thi-Xuan Dang, D.; Barik, R. K.; Phan, M.-H.; Woods, L. M. Enhanced Magnetism in Heterostructures with Transition-Metal Dichalcogenide Monolayers. *J. Phys. Chem. Lett.* **2022**, *13* (38), 8879–8887.

(49) Nguyen, L.-A. T.; Dhakal, K. P.; Lee, Y.; Choi, W.; Nguyen, T. D.; Hong, C.; Luong, D. H.; Kim, Y.-M.; Kim, J.; Lee, M.; Choi, T.; Heinrich, A. J.; Kim, J.-H.; Lee, D.; Duong, D. L.; Lee, Y. H. Spin-Selective Hole–Exciton Coupling in a V-Doped $WSe_2$ Ferromagnetic Semiconductor at Room Temperature. *ACS Nano* **2021**, *15* (12), 20267–20277.

(50) Kresse, G.; Furthmüller, J. Efficient Iterative Schemes for *Ab Initio* Total-Energy Calculations Using a Plane-Wave Basis Set. *Phys. Rev. B* **1996**, *54* (16), 11169–11186.

(51) Kresse, G.; Furthmüller, J. Efficiency of Ab-Initio Total Energy Calculations for Metals and Semiconductors Using a Plane-Wave Basis Set. *Comput. Mater. Sci.* **1996**, *6* (1), 15–50.

(52) Perdew, J. P.; Burke, K.; Ernzerhof, M. Generalized Gradient Approximation Made Simple. *Phys. Rev. Lett.* **1996**, *77* (18), 3865–3868.

(53) Grimme, S.; Ehrlich, S.; Goerigk, L. Effect of the Damping Function in Dispersion Corrected Density Functional Theory. *J. Comput. Chem.* **2011**, *32* (7), 1456–1465.

(54) Dudarev, S. L.; Botton, G. A.; Savrasov, S. Y.; Humphreys, C. J.; Sutton, A. P. Electron-Energy-Loss Spectra and the Structural Stability of Nickel Oxide: An LSDA+U Study. *Phys. Rev. B* **1998**, *57* (3), 1505–1509.

(55) Momma, K.; Izumi, F. *VESTA 3* for Three-Dimensional Visualization of Crystal, Volumetric and Morphology Data. *J. Appl. Crystallogr.* **2011**, *44* (6), 1272–1276.

(56) Wang, V.; Xu, N.; Liu, J.-C.; Tang, G.; Geng, W.-T. VASPKIT: A User-Friendly Interface Facilitating High-Throughput Computing and Analysis Using VASP Code. *Comput. Phys. Commun.* **2021**, *267*, 108033.

# Supporting Information

**Vanadium-doped Molybdenum Disulfide Monolayers with Tunable Electronic and Magnetic Properties: Do Vanadium-Vacancy Pairs Matter?**

Da Zhou[1, 2#], Yen Thi Hai Pham[3#], Diem Thi Xuan Dang[3#], David Sanchez[4], Aaryan Oberoi[5], Ke Wang[6], Andres Fest[4], Alexander Sredenschek[1, 2], Mingzu Liu[1, 2], Humberto Terrones[7], Saptarshi Das[5], Dai-Nam Le[3], Lilia M. Woods[3], Manh-Huong Phan[3,*], and Mauricio Terrones[1, 2, 4, 8,*]

[1]Department of Physics, The Pennsylvania State University, University Park, PA 16802

[2]Center for 2- Dimensional and Layered Materials, The Pennsylvania State University, University Park, PA 16802

[3]Department of Physics, University of South Florida, Tampa, FL 33620

[4]Department of Materials Science and Engineering, The Pennsylvania State University, University Park, PA 16802

[5]Department of Engineering Science and Mechanics, The Pennsylvania State University, University Park, PA 16802

[6]Materials Research Institute, The Pennsylvania State University, University Park, PA 16802

[7]Department of Physics, Applied Physics and Astronomy, Rensselaer Polytechnic Institute, Troy, NY 12180

[8]Department of Chemistry, The Pennsylvania State University, University Park, PA 16802

*Corresponding authors: mut11@psu.edu (M.T.); phanm@usf.edu (M.H.P.)

**Figure S1.** Synthesis scheme, scanning electron microscope (SEM) images, and optical measurements. (a) Schematic figure of the liquid-phase precursor-assisted synthesis. (b) A low-mag SEM image of the flakes, scale bar 200um. (c) High-mag SEM image of a flake, scale bar 10um. (d) Normalized Raman spectra of pristine $MoS_2$, 3 at% $V-MoS_2$, and 8 at% $V-MoS_2$. The spectrum intensities were normalized by each spectrum's silicon peak around 520.5 $cm^{-1}$, which was not included in the plot for view purposes. (e) PL intensities of pristine $MoS_2$ and $V-MoS_2$ of different V concentrations.

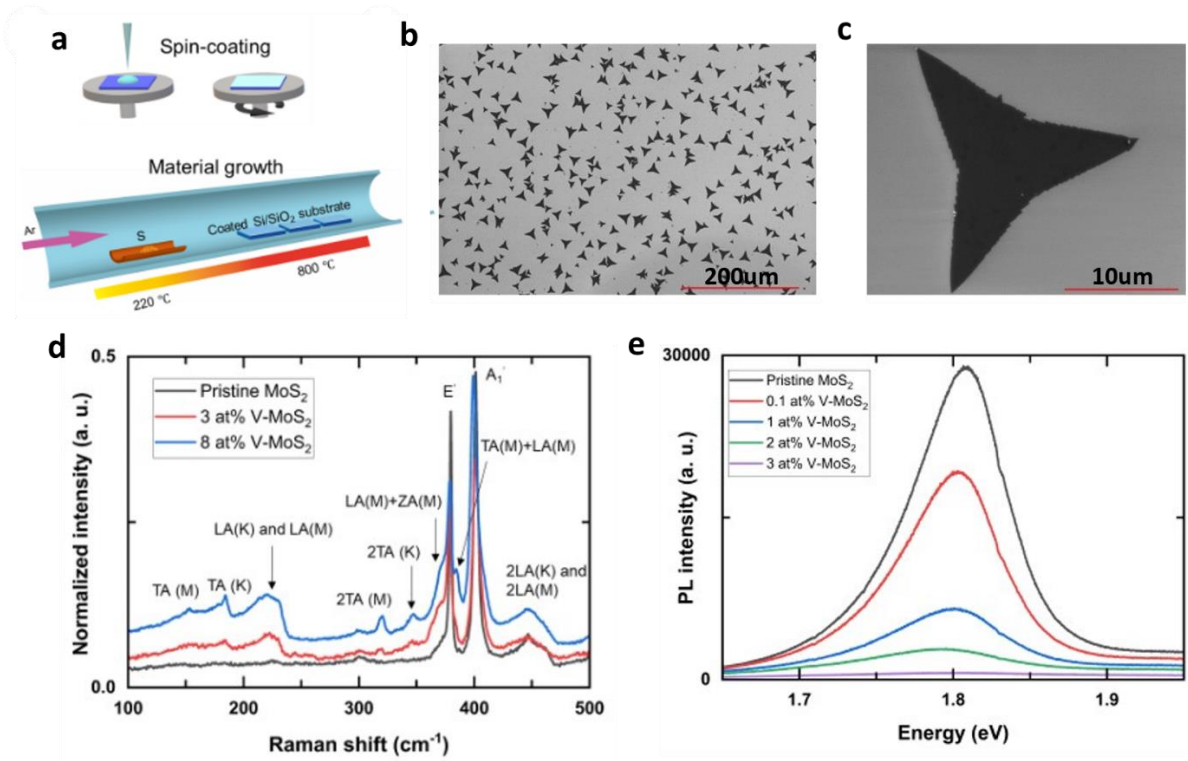

**Figure S2.** XPS spectra of pristine and V-MoS$_2$ monolayers. HAADF-STEM later confirmed that the medium-doped samples are about 3 at.% V concentration, and the high-doped samples are about 8 at.% V concentration.

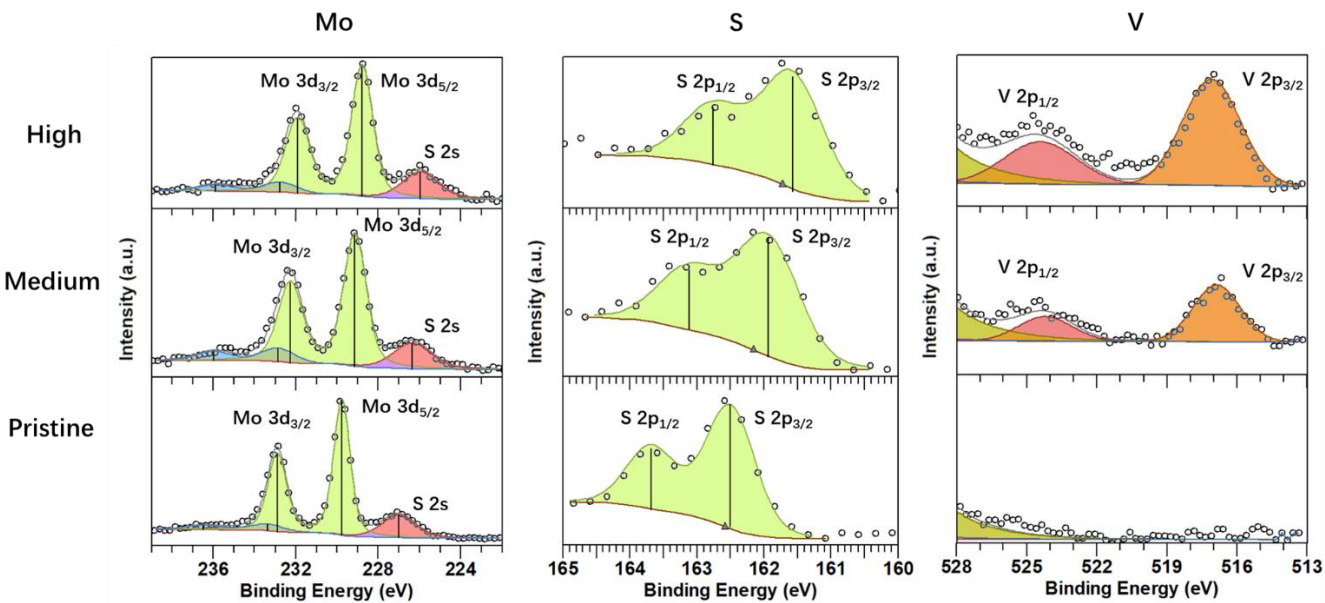

**Figure S3.** HR-STEM images show the two major configurations for the 2 at.% V-MoS$_2$ monolayer: a single V atom coupled with a single S-vacancy (V$_{Mo}$-S$_{vac}$) and single V atom coupled with two nearest S-vacancies (V$_{Mo}$-2S$_{vac}$). Yellow circles represent V atoms, while red circles represent S vacancies.

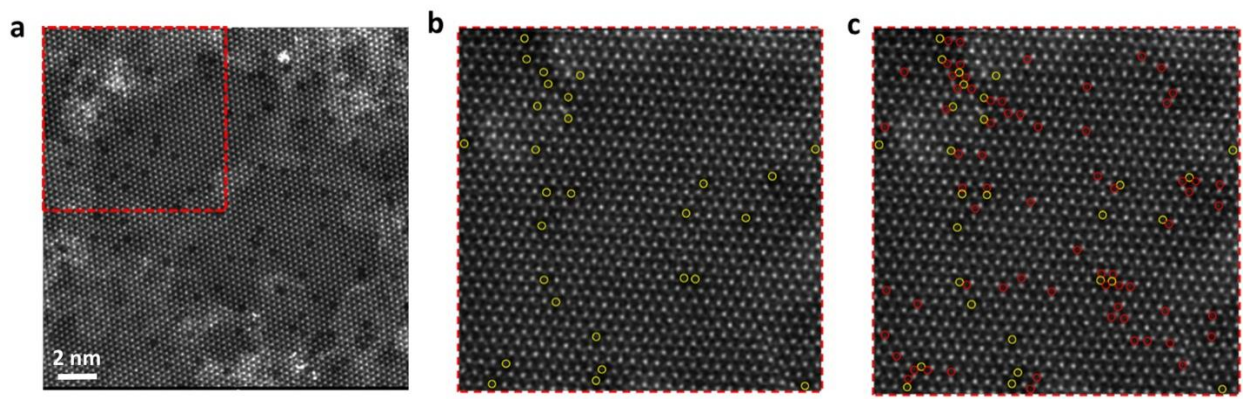

**Figure S4.** Magnetic loop *M*(*H*) of the 8 at.% V-MoS$_2$ monolayer sample at room temperature. The monolayer shows no magnetic ordering, indicating a complete suppression of ferromagnetism when the high V concentration (8 at.%) is introduced to the MoS$_2$ lattice.

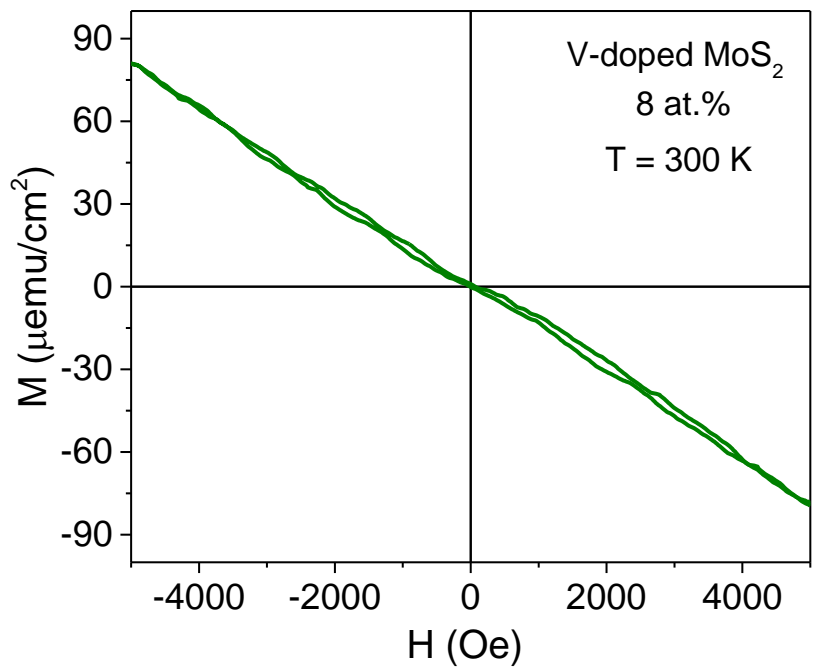

**Figure S5.** Magnetic hysteresis loops *M*(*H*) taken at 10 K under zero-field-cooled (ZFC) and field-cooled (FC) measurement protocols for the 3 at.% V-MoS$_2$ monolayer sample. It is worth noting that no shift in the FC *M*(*H*) loop with reference to the ZFC *M*(*H*) loop has been observed in the 3 at.% V-MoS$_2$ monolayer. This points to the dominance of V-V FM interactions (rather than AFM interactions due to S/Mo vacancies or dopant-vacancy pairs) in this sample.

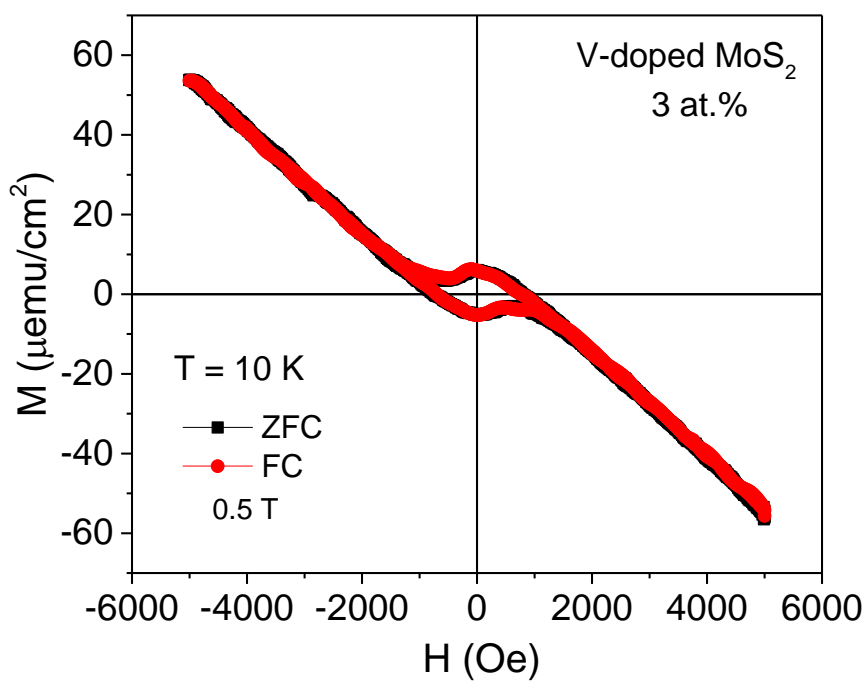

**Figure S6.** Temperature dependence of magnetization measured in a field of 0.5 T under zero-field-cooled (ZFC) and field-cooled (FC) measurement protocols for the 3 at.% V-MoS$_2$ monolayer sample.

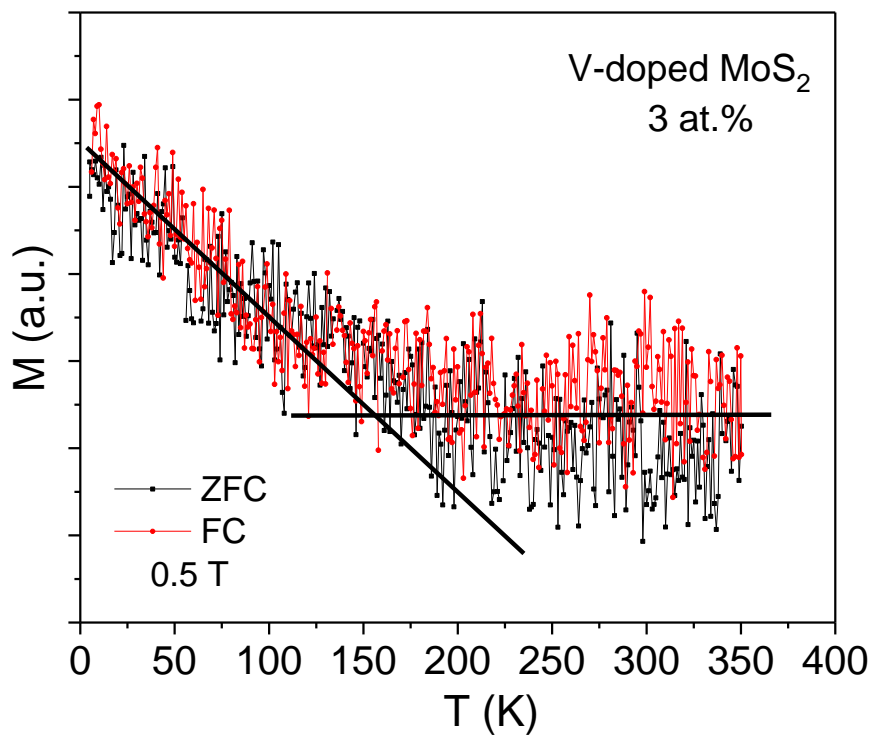